\documentclass[11pt,a4paper]{article}
\usepackage[makeroom]{cancel}
\usepackage{bm}
\usepackage{amsmath}
\usepackage{amsfonts}
\usepackage{graphicx}
\usepackage{nicefrac}
\usepackage{authblk}
\usepackage{cite}
\usepackage[left=2.2cm,top=2.2cm,right=2.2cm,bottom=2.2cm,nohead]{geometry}
\DeclareGraphicsExtensions{.png,.jpg,.pdf}
\usepackage{epstopdf}
\usepackage{float}
\usepackage{subfigure}
\usepackage{fouriernc} 
\usepackage{enumitem}
\setlist{nolistsep,leftmargin=*}
\usepackage{mdframed}
\usepackage[version=4]{mhchem}
\usepackage{soul}

\usepackage{setspace} 
\usepackage{etoolbox}
\AtBeginEnvironment{quote}{\singlespacing\small}
\usepackage{xcolor}
\definecolor{myblue}{RGB}{212,239,253}

\title{The difference between Faradaic \\ and non-Faradaic electrode processes} 

\usepackage{authblk}
\makeatletter
\renewcommand\AB@authnote[1]{\textsuperscript{\normalfont#1}}

\makeatother

\author[1]{P.M.~Biesheuvel,}
\author[1]{S.~Porada,}
\author[2]{J.E.~Dykstra}

\affil[1]{Wetsus, European Centre of Excellence for Sustainable Water Technology, 
The Netherlands.}
\affil[2]{Environmental Technology, Wageningen University, The Netherlands.}

\date{} 

\begin{document}


\renewcommand{\t}{\widetilde}
\renewcommand{\t}{}
\newcommand{\s}[1]{\mathrm{_{#1}}}

\maketitle

\begin{abstract}
Both Faradaic and non-Faradaic processes can take place at an electrode. The difference between the two processes is clearly discussed in several classical sources, starting with Grahame (1952). However, later reference to charge transfer across the metal-solution interface as a defining feature of a Faradaic process, has led to ambiguities. Following Grahame, in a Faradaic process, charged particles transfer across the electrode, from one bulk phase to another. Thus, in a Faradaic process, after applying a constant current, the electrode charge, voltage and composition go to constant values. Instead, in a non-Faradaic (capacitive) process, charge is progressively stored. We characterize the intercalation material nickel hexacyanoferrate by two electrochemical methods and compare with theory. Data for the capacitance of this material is well described by the extended Frumkin isotherm. This data, and the correspondence with theory, demonstrates that this is a capacitive material and ion and charge storage in this material a non-Faradaic electrode process. Cyclic Voltammetry (CV) diagrams for this material have broad peaks for certain potential windows, and rectangular shapes for other conditions, both experimentally and in theoretical calculations based on a RC network model that includes how capacitance is a function of charge. Measured and predicted CV diagrams are in perfect agreement with one another. This shows that (broad) peaks in CV diagrams do not establish whether an electrode material is Faradaic or not. 
\end{abstract}

\begin{center}{\noindent\rule{12cm}{0.4pt}}\end{center}

\begin{center}


\end{center}

\medskip

In electrochemistry and electrochemical engineering there is a clear 
distinction between two types of processes that can take place in an electrode. These are called Faradaic and non-Faradaic processes~\cite{Grahame,Vetter,Mohilner,Parsons,Erdey-Gruz,Bard_Faulkner_1980}. 
This difference is important to make because the two processes represent two fundamentally different modes of how an electrode operates, and the distinction determines how an electrode (process) can be characterized experimentally. 
For a Faradaic process we can construct a steady-state current-voltage (\textit{i}-\textit{V}) curve, often called a polarization curve, and this makes it categorically different from a non-Faradaic process, in which case an (equilibrium) charge-voltage ($\sigma$-\textit{V}) curve can be constructed. Thus, the proper class of an electrode process can 
be established on the basis of experiments using nothing more than an electrometer (a high impedance digital voltmeter commonly used to measure electrode potentials) without detailed knowledge of atomistic details of the electrode process. 


\newpage

The difference between a Faradaic and non-Faradaic electrode process was already clearly explained by David Grahame in 1952: 

\begin{center}\textit{There are two ways in which the current is carried across the interface of a metal-electrolyte system, and these two may be called the faradaic and nonfaradaic paths, respectively. In the former, current crosses the interface by virtue of an electrochemical reaction such as the reduction or oxidation of water or of some ion. In the latter (nonfaradaic) case charged particles do not cross the interface, and the current is carried by the charging and discharging of the electrical double layer, which behaves like a condenser in series with the Ohmic resistance of the solution.}\textsuperscript{3} 

\textit{\textsuperscript{3}A possible ambiguity needs to be considered here. How is one to differentiate between a faradaic and a nonfaradaic current? The answer is that any process which allows a continuous current to flow will be regarded as faradaic, whereas one which does not will be regarded as nonfaradaic. [...] }

\textit{The question of whether or not a continuous current flows hinges upon the question of whether or not the products of electrolysis can build up in concentration (or more strictly in chemical potential) 
in such a manner as to stop the flow of current. If one or more of the products of electrolysis can diffuse away, this will never happen, since more current will be needed to replace the substance which has diffused away.} \end{center}

\begin{figure}[H]
\centering
\includegraphics[width=0.8\textwidth]{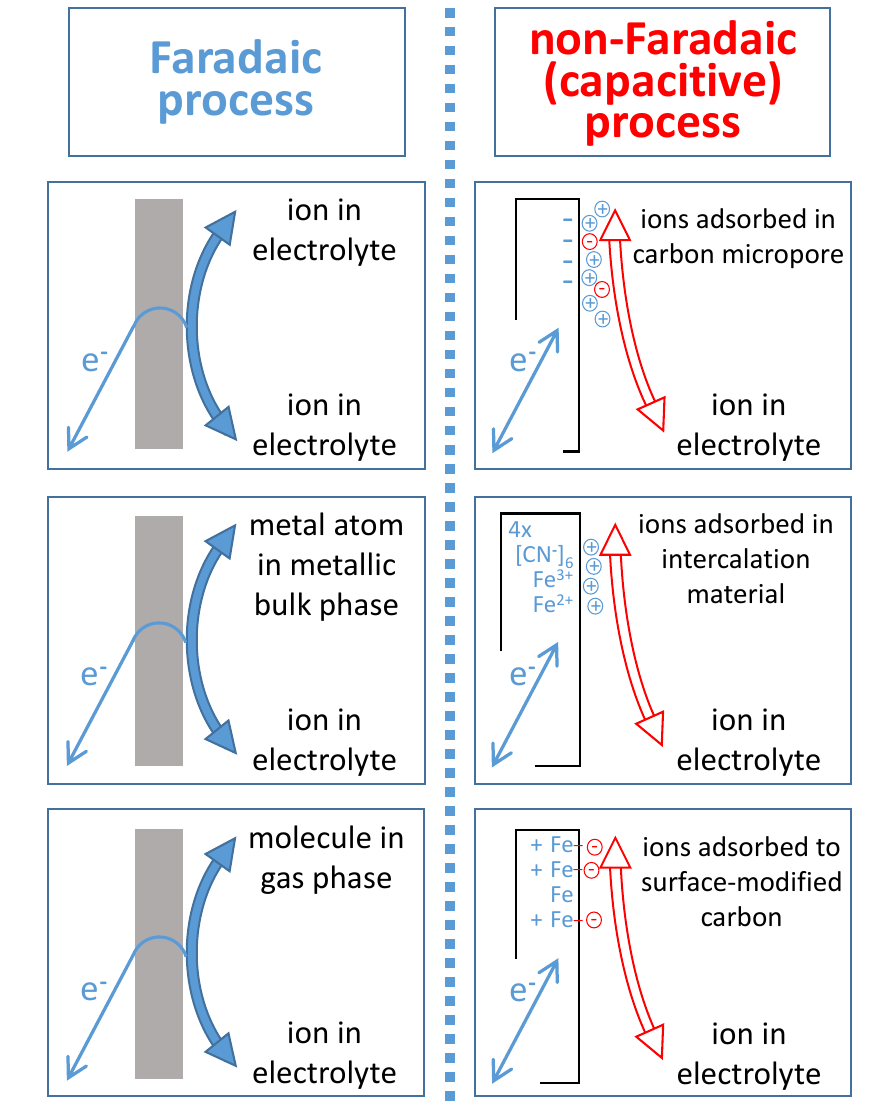}
\vspace{-1pt}
\caption{The storage of ions and charge as 
the key difference between Faradaic and non-Faradaic (capacitive) electrode processes, using six examples.}
\label{fig_cartoon}
\end{figure}

The complete text by Grahame on this topic is provided below. Thus, to have a Faradaic process, there must be an electrode reaction where an ion or atom is reduced or oxidized to another species, and in addition 
the reactant must 
come from outside the electrode, and the product of the reaction must leave the electrode again, see Fig.~\ref{fig_cartoon}. The outside phase where the product goes to, can be the same bulk phase as where the reactant came from, or another bulk phase. Bulk phases are the phases outside the electrode, and the electrode is the interface (or alternatively called `interphase') 
between such bulk phases, of which at least one part 
conducts electronic charge, and at least one part 
allows for ion transport. Thus, in a Faradaic process, reactants and products of the electrode reaction enter and leave the electrode.

In a Faradaic process, in most cases there is the transfer of electrons across the electrode, from a conducting (metallic) bulk phase, to the electrolyte, or vice-versa. If electrons move to the electrolyte, the ions (or other molecules) that enter the electrode in reduced form leave as oxidized species when they return to the electrolyte phase (or alternatively to another bulk phase, such as a gas phase or solid salt). 
In other Faradaic processes, it is the ion that crosses the electrode, coming from an electrolyte phase and moving to the bulk metal phase that supplies the electronic charge, which is what happens in metal plating. (Or the reverse, an atom dissolves as an ion.) 

In both these types of Faradaic processes, charged particles (electrons or ions) transfer across the electrode, i.e., they enter and they leave, coming from one bulk phase, and either returning to the same bulk phase, or going to another one. In neither case is 
charge progressively stored in the electrode. 
%
%
One of the bulk phases in contact with the electrode will be an electrolyte phase (either liquid or solid), in which ions or uncharged molecules are dissolved (or are a constituent of the solvent), but the other bulk phase 
can also be something else, such as a bulk metal phase, as is the case for a plating reaction. Alternatively, an oxidic layer formed on a metal can serve as the bulk phase, 
or a solid salt, as is the case for a corrosion process, for the lead acid battery, or for the Ag/AgCl electrode. Finally, one of the bulk phases can also be a gas phase, as for hydrogen fuel cells. 

With ionic (atomic, molecular) reactants coming from such a bulk phase, and products eventually going there, we 
have a Faradaic process. In this case we can characterize the electrode process by construction of a polarization curve (\textit{i}-\textit{V} curve). This works just as well when the electrode reaction is part of a chain of reaction steps, where 
prior to and after the electrode reaction the ions (atoms, molecules, adsorbed species) are involved in transport and reaction steps 
such as 
adsorption and desorption, and association or dissociation reactions on the metal surface, all without an electron yet reacting with an atomic species. Also in these more complicated reaction sequences, the key requirement for a process to be Faradaic remains that the electrons and ions that enter the electrode and undergo an electrode reaction, also leave the electrode again and as a product species go to an external bulk phase. For such a Faradaic process, as long as the external bulk phases do not change their composition (and do not disappear altogether), steady-state operation can be established, where each value of the current corresponds to one value of the electrode potential, and vice-versa. To some extent 
the electrode merely serves as a `meeting place' between on the one hand electronic charge, and on the other hand molecular, atomic or ionic species from adjoining bulk phases. 
Because the products of the electrode reaction leave the electrode, 
the Faradaic process can continue forever, as long as the current keeps on flowing and the bulk phases can supply reactants and take up products.

In contrast, in a non-Faradaic electrode process, when current flows, we will notice that charge is progressively stored. This is because ions or other species that enter the electrode cannot leave. This is the case when there is no electrode reaction at all, or the electrode reaction involves atoms that are part of the electrode structure itself, or when the reacting species is oxidized/reduced in the electrode, but then stays there, i.e., `is locked in the electrode'. For some reason the chain of reaction steps in the electrode (as described above) is truncated at some point. 
In these cases there is no transfer of ions or electrons across the electrode (from one bulk phase to another bulk phase). 
In all of these situations we can construct a $\sigma$-\textit{V} curve, and not an \textit{i}-\textit{V} curve. Examples are electrodes consisting of an intercalation material such as nickel hexacyanoferrate, a Prussian Blue Analogue, where the $\ce{Fe^2+}$/$\ce{Fe^3+}$ lattice atoms can be oxidized and reduced~\cite{Porada_Smith}; or the ion storage in an electrode that has redox-active ferrocene group immobilized on a carbon surface~\cite{Su_Hatton}; or when \ce{Li^+}-ions in organic solvent that transfer into an intercalation electrode 
are (partially) reduced to metallic \ce{Li}-atoms which stay in the electrode. The $\sigma$-\textit{V} curve of these electrode materials can be analyzed to derive values for capacity (a number typically with unit C/g), or capacitance (F/g), see Fig.~\ref{fig_capacitance} for an example. This is why non-Faradaic electrode processes can also be called capacitive processes.

Next, we provide verbatim texts from six classical texts that discuss these two kinds of processes in detail. The first is the aforementioned 1952 paper by D.C.~Grahame and the other five are books or book chapters, by K.J.~Vetter (1961/1967), D.M.~Mohilner (1966), Parsons (1970), T.~Erdey-Gr{\'u}z (1972), and A.J.~Bard and L.R.~Faulkner (1980)~\cite{Grahame, Vetter, Mohilner,Parsons,Erdey-Gruz,Bard_Faulkner_1980}. We also provide their words on nonpolarizable and ideally polarizable electrodes, as these terms line up exactly with the difference between a Faradaic and non-Faradaic process. [All italicization is from the original. Underlining in the Grahame text is ours.]

\begin{figure}[H]
\centering
\includegraphics[width=0.8\textwidth]{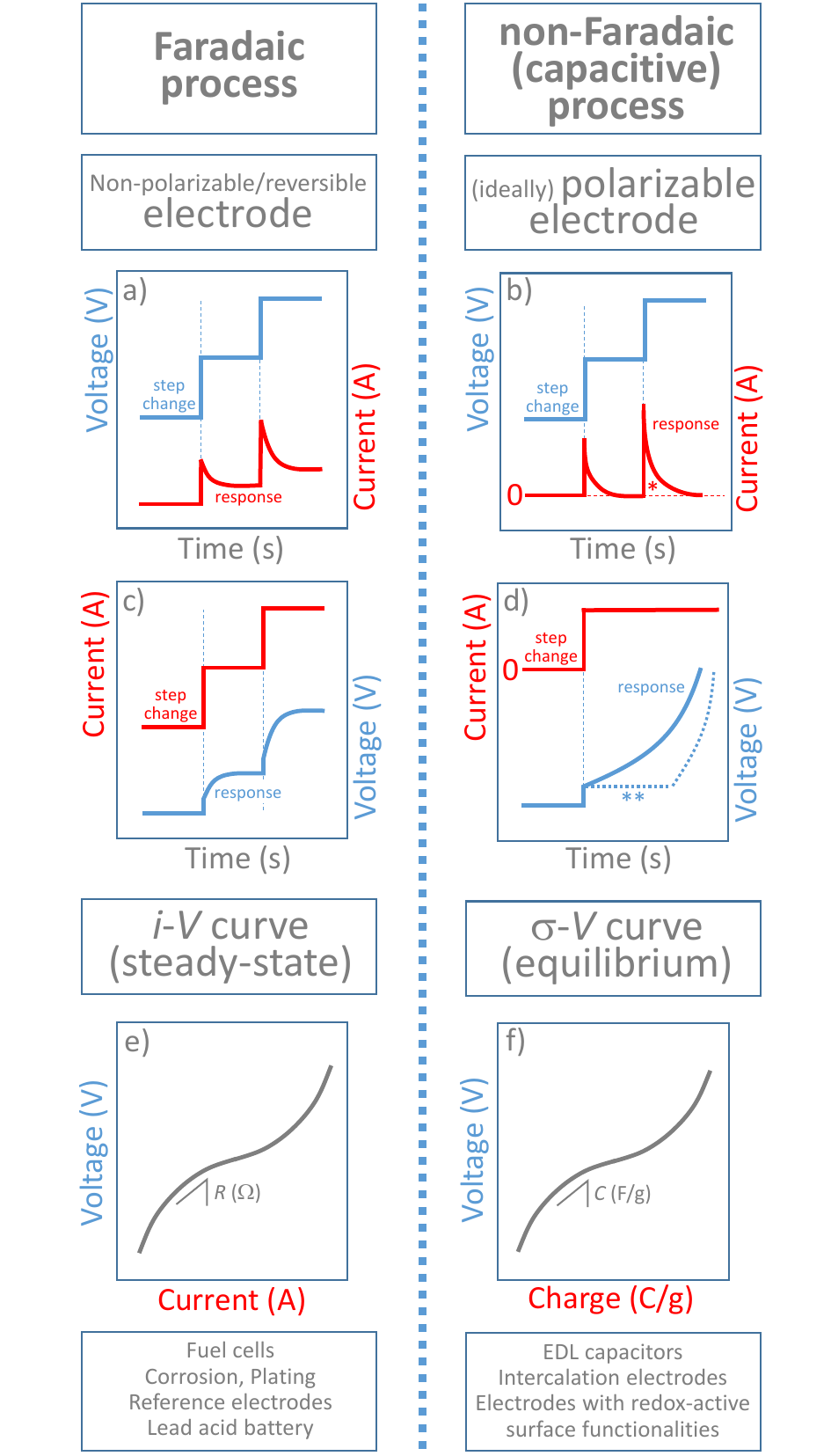}
\vspace{-1pt}
\caption{Operational differences between Faradaic and non-Faradaic (capacitive) electrode processes.}
\label{fig_overview}
\end{figure}



\noindent \textbf{Grahame (1952)}

\begin{quote}
\begin{mdframed}[backgroundcolor=myblue,linecolor=white] 

There are two ways in which the current is carried across the interface of a metal-electrolyte system, and these two may be called the faradaic and nonfaradaic paths, respectively. In the former, current crosses the interface by virtue of an electrochemical reaction such as the reduction or oxidation of water or of some ion. In the latter (nonfaradaic) case charged particles do not cross the interface, and the current is carried by the charging and discharging of the electrical double layer, which behaves like a condenser in series with the Ohmic resistance of the solution.\textsuperscript{3} 

\textsuperscript{3}A possible ambiguity needs to be considered here. How is one to differentiate between a faradaic and a nonfaradaic current? \ul{The answer is that any process which allows a continuous current to flow will be regarded as faradaic, whereas one which does not will be regarded as nonfaradaic}. [...] 
The question of whether or not a continuous current flows hinges upon the question of whether or not the products of electrolysis can build up in concentration (or more strictly in chemical potential) 
in such a manner as to stop the flow of current. If one or more of the products of electrolysis can diffuse away, this will never happen, since more current will be needed to replace the substance which has diffused away. 

Likewise if the product of electrochemical action is capable of undergoing a second reaction by which its concentration (chemical potential) is lowered, a faradaic current will flow to replace the substance used up in the chemical reaction. (This happens when hydrogen atoms combine to form hydrogen molecules.) Finally, a faradaic current will flow if the product of reaction has a natural limit of chemical potential which is reached before the counterelectromotive force needed to stop the reaction is attained. This happens very often in electrochemical processes, as in the deposition of a metal or the formation of an insoluble salt with the metal of the electrode. The evolution of a gas at an electrode is usually not an example of this effect, however, since the production of gaseous molecules does not ordinarily occur in a single electrochemical step.

This discussion also bears upon the question of the distinction between an adsorbed ion and an ion which has acquired or lost electrons to become an atom. Even the adsorbed ion is associated with an equal amount of the opposite charge in the double layer, so that the distinction cannot be made simply on the basis of the charge. \ul{The distinction is made on the basis of whether or not the adsorption is such as to allow a continuous current to flow}, as discussed above. Thus if the ``adsorbed'' ion together with its associated charge can diffuse away from the interface as a unit, the ion has really reacted with the charge. Likewise, if the adsorbed ion has formed a soluble or insoluble salt with the metal, thereby exposing more metallic surface and allowing more current to flow, the ion has in fact reacted and is not to be considered adsorbed. 

\end{mdframed}
\end{quote}

(Underlining not in original.)
\medskip


\noindent \textbf{Vetter (1961, 1967)}

\begin{quote}
\begin{mdframed}[backgroundcolor=myblue,linecolor=white] 

... is a reaction in which charge carriers (ions and electrons) are transferred across the electrical double layer at a phase boundary; such reactions are therefore called charge-transfer reactions. 


... As before, [current] \textit{i} will continue to be only the charge-transfer current density (faradaic current) in contrast to the capacitive current density $i_C = C_D \cdot d\varepsilon / d t$ which leads to the charging of the double-layer capacitance $C_D$. Together both yield the total current density $i_* = i + i_C$. 

\end{mdframed}
\end{quote}


\medskip


\noindent \textbf{Mohilner (1966)}

\begin{quote}
\begin{mdframed}[backgroundcolor=myblue,linecolor=white] 

\textbf{Ideal Polarized Electrodes}. Ideal polarized electrodes [...] are defined as electrodes at which no charge transfer across the metal-solution interface can occur, regardless of the potential imposed on the electrode from an outside source of voltage. 

In a given solution at any fixed potential within the permissible range, the double layer at an ideal polarized electrode attains a true state of equilibrium which can be described precisely in terms of classical equilibrium thermodynamics. However, this equilibrium is not of the familiar nerstian type. Rather, it is a state of \textit{electrostatic equilibrium} in the electrical double layer. Therefore, to define the state of an ideal polarized electrode at equilibrium, it is necessary to specify not only the temperature, pressure, and composition (chemical potentials) of each phase, but also the value of an additional electrical variable. This electrical variable expresses the degree of charge separation across the interface. Depending on convenience, one may choose for the electrical variable either the excess charge density on the metal $q^M$ or the potential $E$ of the ideal polarized electrode with respect to a reference electrode. Thus, an ideal polarized electrode at equilibrium is a system having one more degree of freedom than it would have were it in a state of nernstian equilibrium. This means that an ideal polarized electrode has the unique capability of being in thermodynamic equilibrium at any potential whatever (within a certain range), although the temperature, pressure, and composition of its phases remain fixed.

\textbf{Charge-Transfer Electrodes}. Electrodes which are not ideal polarized may be called \textit{charge-transfer electrodes}. At these electrodes the familiar electrochemical processes of oxidation and reduction take place. In terms of the electrical double layer, a charge-transfer electrode is one at which electrically charged particles, ions or electrons, can be transferred across the metal-solution interface. In electrical terminology, a \textit{conduction current} can flow across the interface of a charge-transfer electrode, but only a \textit{displacement current} can flow at the interface of an ideal polarized electrode.

For fixed temperature, pressure, and composition of each phase, there is one, and only one, value of the electrode potential for which a charge-transfer electrode may be at equilibrium. This is the potential specified by Nernst's equation. In contrast, for an ideal polarized electrode to be at equilibrium under the same conditions, any of a continuously infinite set of potentials will suffice.

\textbf{Faradaic and Nonfaradaic Processes}. The familiar electrode processes of oxidation and reduction which take place at charge-transfer electrodes obey Faraday's laws; hence they are called \textit{faradaic}. At an ideal polarized electrode, faradaic processes are prohibited. Whenever a real electrode behaves as an ideal polarized electrode, it is because, within a certain range of potentials, all the faradaic processes which might conceivably take place there fall into either of two categories; (a) The activation energy is so high that the faradaic process occurs at a negligible rate. [...] (b) Even though the activation energy is low, the equilibrium constant for the faradaic process is such that the concentration of either reactants or products is so low as to be meaningless (except in a statistical sense). Therefore, any charge transfer accompanying a change of electrode potential is entirely negligible. 

The processes of adsorption and desorption which take place whenever the structure of the electrical double layer changes are not described by Faraday's law; hence they are called nonfaradaic. At ideal polarized electrodes, only nonfaradaic processes can take place, but at charge-transfer electrodes, both faradaic and nonfaradaic processes occur simultaneously.

\end{mdframed}
\end{quote}

\medskip


\vspace*{\fill}

\noindent \textbf{Parsons (1970)}

\begin{quote}
\begin{mdframed}[backgroundcolor=myblue,linecolor=white]

Faradaic processes are defined as those which obey Faraday's law, that is the amount of chemical reaction occurring 
is directly proportional to the amount of charge passed across the electrode boundary. When the system is in a steady-state the application of this definition is simple: all the current is faradaic. Difficulties begin with the study of transient processes, because of the formation of adsorbed species whose concentration is time dependent. The adsorption of charged species (or of uncharged species formed from charged species) is equivalent to the storage of charge at the interface and hence to a nonfaradaic contribution to the observed current.

\end{mdframed}
\end{quote}

\medskip

\vspace*{\fill}

\noindent \textbf{Erdey-Gr{\'u}z (1972)}

\begin{quote}
\begin{mdframed}[backgroundcolor=myblue,linecolor=white] 

\textbf{Polarization and overvoltage}. Usually the passage of electric current changes the potential of the electrode, a phenomenon called \textit{polarization}.  

Polarization brings about changes in the electrochemical double layer at the electrode surface. As a first approximation, the double layer can be regarded as a capacitor. If ion formation and neutralization are slow, the ions reaching the double layer as a result of current flow tend to increase the charge of the capacitor and the potential between its plates. This fraction of the current, associated with all polarizations, is the so-called \textit{capacitive current} which stops as soon as the steady state is reached. Since charge transfer always occurs at the electrode, even if it is slow, the double layer can be regarded as a capacitor with some leakage due to a parallel resistance. The fraction of current which is not involved in changing the charge of the double-layer capacitor but passes across the phase boundary by electron transfer has come to be known as \textit{faradaic current}. 

If the faradaic current is disregarded, the amount of charge carried by current $i$ (A cm$^{-2}$) during a time interval $\text{d}t$ into the double layer is given by $\text{d}q=i \text{d}t$. This causes the potential difference across the double layer to change [...].  

\textbf{Depolarization; Polarizable and non-polarizable electrodes}. Upon transferring electric charge to electrodes by means of charge-carrier ions, various changes occur. The charge carriers (ions, electrons) reaching the surface of an indifferent electrode (i.e. one which does not release nor neutralize ions) cannot undergo discharge to form components of neutral particles. Under such conditions, the charge carriers enter the double-layer capacitor located at the phase interface and change the amount of charge on, and the potential difference across, its plates. The change of the potential difference is reflected by the polarization of the electrode. Systems which behave in this manner are \textit{ideally polarizable electrodes}. 

By means of ideally polarizable electrodes, the charge and potential of the electrochemical double layer can be varied freely within certain limits. The variation of these parameters permits the study of the structure of the double-layer. 

There is no thermodynamic equilibrium between an ideally polarizable electrode and the solution because there is no common component capable of changing its charge and being transferred between the phases, conditions necessary for equilibrium. The state of an ideally polarizable electrode is well defined only if an external source is used to maintain a constant polarization potential, i.e. the double-layer capacitor charged with a definite charge. The polarization potential is an independent parameter of the system. 

Ideal polarizability can only be realized, even approximately, in a limited potential range. In all cases, if the potential becomes sufficiently positive or negative, some electrode processes will start to occur, i.e., charge will be transferred between the plates of the capacitor. If no other process can take place, hydrogen or hydroxide ions will be neutralized from aqueous solutions. As a result, the charge on the plates, and the potential difference between them, will be decreased, i.e. \textit{depolarization} occurs. 

[If] depolarization is strong [...] and the process is fast enough, then metal and solution are in thermodynamic equilibrium. Under such conditions the electrode potential is but slightly changed by current flowing through the electrode. The reason is that changes of charge and potential in the double-layer due to the flow of current accelerate electron transfer. Thus the process becomes fast enough to compensate for any changes of electron concentration on the surface before such changes would cause an appreciable potential shift in the double-layer. The potential of electrodes characterized by large exchange currents is practically unaffected by small current densities (\textit{non-polarizable electrodes}), and large current densities only affect the potential in so far as the concentration of the potential-determining ions changes around the electrode [...]. 

\end{mdframed}
\end{quote}

\vspace*{\fill}

\medskip


\noindent \textbf{Bard and Faulkner (1980)}

\begin{quote}
\begin{mdframed}[backgroundcolor=myblue,linecolor=white] 
\textbf{Faradaic and Nonfaradaic processes}. Two types of processes occur at electrodes. One kind comprises those just discussed, in which charges (e.g., electrons) are transferred across the metal-solution interface. This electron transfer causes oxidation or reduction to occur. Since these reactions are governed by Faraday's law (i.e., the amount of chemical reaction caused by the flow of current is proportional to the amount of electricity passed), they are called \textit{faradaic} processes. Electrodes at which faradaic processes occur are sometimes called \textit{charge transfer} electrodes. 

Under some conditions a given electrode-solution interface will show a range of potentials where no charge transfer reactions occur because such reactions are thermodynamically or kinetically unfavorable. However, processes such as adsorption and desorption can occur, and the structure of the electrode-solution interface can change with changing potential or solution composition. These processes are called \textit{nonfaradaic} processes. Although charge does not cross the interface under these conditions, external currents can flow (at least transiently) when the potential, electrode area, or solution composition changes. Both faradaic and nonfaradaic processes occur when electrode reactions take place. 
        
\textbf{Nonfaradaic processes and the nature of the electrode-solution interface.}

\textbf{The ideal polarized electrode}. An electrode at which no charge transfer across the metal-solution interface can occur regardless of the potental imposed by an outside source of voltage is called an \textit{ideal polarized} (or \textit{ideal polarizable} \footnote{From Bard and Faulkner, 2001}) \textit{electrode} (IPE). While no real electrode can behave as an IPE over the whole potential range available in a solution, some electrode-solution systems, over certain limited potential ranges, can approach ideal polarizability. 

\textbf{Capacitance and Charge of an Electrode}. Since charge cannot cross the IPE interface when the potential across it is changed, the behavior of the electrode-solution interface is analogous to that of a capacitor. 
\end{mdframed}
\end{quote}

\bigskip

These six sources 
provide much useful background information on the different electrode processes. They take a similar perspective in their focus on what can be measured about electrode behavior, basically using an electrometer only, and how those observations are to be interpreted. However, in some of these texts 
an ambiguity arises, 
because to distinguish Faradaic from non-Faradaic processes 
the concept of `transfer of charge across the metal-solution interface' is introduced. 
However, there are two ambiguous aspects to this concept, and we discuss them in the next two paragraphs.

The problem with the term `charge transfer' is that 
in all electrode processes, steady state and dynamic, capacitive and Faradaic, there is always perfect charge transfer: the electronic current arriving in the electrode is exactly equal to the ionic current leaving, with the electrode as a whole remaining electroneutral. 
Instead, the defining feature of a Faradaic process is not that \textit{charge} is transferred across the electrode (because that always happens), but \textit{charged particles} (electrons or ions) are transferred across the electrode, from one bulk phase to another (with bulk phases being external to the electrode), in line with the wording of Grahame (1952). 
Thus how we see it, is that in a Faradaic process there is transfer across the electrode of electrons or ions (or other material species), and neither accumulates in the electrode.  
Thus, the term `charge transfer electrode' is only a term indicative of a Faradaic process if it is implied that it is \textit{charged particles} (electrons, ions or other species) that transfer across the electrode, from one to another bulk phase, not staying behind, not being stored in the electrode. 

The other problem is the `across the metal-solution interface'-part. 
This may be mistaken to refer to 
the idea that we must 
speak of 
a Faradaic electrode if there is a degree of mobility 
of an ion or electron across some theoretical surface (a 2D infinitely thin plane) 
\textit{inside the electrode}. 
However, the invocation 
of such a theoretical surface to distinguish two classes of processes is problematic, because it is an element of a theoretical representation of an electrode. 
And thus many such planes can be assumed, so which plane to choose? 
And how to be sure there was transfer of charged particles across it? 
Thus any conclusion that is based on such a theoretical approach 
depends strongly on the researcher's view of 
the inner workings of the electrode, i.e., on the theory 
of the atomistic details of what may or may not occur inside the electrode. For instance it depends on one's 
view of whether or not an electron moves out of the metallic region 
into the ionic region 
--all within the electrode-- and there 
associates 
with ions, or whether the electron stays in the metallic structure, with the ion residing nearby. But that microscopic perspective 
is always up for discussion, and the validity of any such atomistic model can always be questioned at a later time, and the preferred theory 
can change. And thus a process that is considered Faradaic at one time can become non-Faradaic at a later time, and vice-versa. This is 
a sub-optimal situation. It is preferable when the designation of a Faradaic process versus a non-Faradaic (capacitive) process is more robust 
and does 
not hinge on the choice of model of what happens on the atomic scale in an electrode. The robust definition is based on how 
the electrode functions in a process, and this operational behaviour can be probed experimentally. 

Indeed, the atomistic or microscopic perspective just discussed, it neglects the observational or phenomenological side of the study of electrodes, of what can be measured about the response of currents and voltages in an experiment that basically only requires an electrometer. This latter perspective is also predominantly taken in the texts cited above. 
Especially interesting is the last paragraph of the full text by Grahame (1952),  
which reads like an early identification of exactly the problem inherent in the microscopic approach, which is that it is unknown what might or might not happen inside the electrode. He (therefore) preferred the experimental/phenomenological approach, and focused on the experimentally accessible information of whether or not the electrode changes its composition (stores both electronic and ionic charge) 
upon ongoing current supply. 
In his view, and we share that view, this is what distinguishes 
a Faradaic process from a non-Faradaic (capacitive) process, and this distinction can be clearly established by simple experiments, with the conclusions independent of the microscopic model one puts forward about what happens inside the electrode. Making the distinction in this way is very helpful because for a Faradaic process very different metrics and characterization methods apply than for a non-Faradaic (capacitive) process. Instead, if one 
focuses strongly on the microscopic perspective and atomistic models, one often 
arrives at a discrepancy between vocabulary derived from the proposed atomistic model (e.g., the electrode is Faradaic), and the observational/experimental behaviour of an electrode (e.g., that it behaves capacitively), and this then leads to much additional vocabulary to resolve 
this mismatch of perspectives, 
describing how a material is in the one class, but behaves as if it is in the other class~\cite{Costentin}.

Thus, microscopic considerations about whether or not there is charge transfer across a certain plane within the electrode, with an electrode reaction possibly taking place there, 
are not very helpful in distinguishing between a Faradaic and non-Faradaic process. Instead, the phenomenological perspective 
clearly 
distinguishes between the two classes of electrode process, a 
perspective also taken in the classical sources. 
The defining criterion is whether or not the electrode changes composition upon steady current supply, see Fig.~\ref{fig_overview}. If the electrode composition does not change, we have a 
nonpolarizable electrode, and the process is Faradaic. In a Faradaic process, reactants and products of the electrode reaction (via intermediate processes at the electrode), ultimately come from, and end up in, a bulk phase, such as a solid metal, solid salt (layer), electrolyte, or gas phase. In contrast, for an ideally polarizable electrode, and thus a non-Faradaic process, there either is no electrode reaction in the electrode at all, or the reaction 
involves an atom, molecule or group that stays bound inside the electrode. As a consequence, for such a non-Faradaic process there really is storage of ions (ionic charge) and electrons (electronic charge), 
and thus 
the overall composition of the electrode changes upon ongoing current supply,  
which will be reflected in a changing electrode potential. In the Faradaic process this is not the case, the electrode does not store ionic and electronic charge, 
and applying a steady current will not change the composition of the electrode over time. 

With reference to Fig.~\ref{fig_overview}, let us reiterate the differences between Faradaic and non-Faradaic processes once again. 
This difference also 
exactly lines up with that between 
non-polarizable 
and 
(ideally) polarizable electrodes. At several points this is also implied in the source texts. 
We can distinguish between the two processes based on how they respond to a step change in voltage (panels a and b) or current (panels c and d). Upon a step change in electrode potential (panels a and b), the Faradaic process quickly levels out to a new value of the current (different from before). Instead, in the non-Faradaic process, after a voltage step change, after some time the current will return back to zero, and it will do so after each step in voltage. The integral of current with time, denoted by~* in panel b, is the additionally stored charge. Upon applying a current step (panels c and d), the Faradaic process responds by going to a new electrode potential, while in the non-Faradaic process, any ongoing nonzero current will either result in the voltage increasing without limit, or the voltage is constant for a limited period of time 
before it also starts to increase~\cite{Dreyer}. This second scenario is possible for certain types of battery electrodes with phase separation of the ions inside the electrode 
In this case, the width of the plateau (** in Fig.~\ref{fig_overview}d) is proportional to the amount of electrode material tested. 
Two or more voltage plateaus are possible when more than two phases form~
\cite{KinnonHearing,Whittingham,Jacobsen}. 

Based on data from such experiments, we can construct two types of defining characteristic curves. For the Faradaic process, we can construct a current-voltage curve based on steady-state data, the \textit{i}-\textit{V} curve, or polarization curve, see Fig.~\ref{fig_overview}e. The slope of the curve has the unit of Ohm (or e.g. $\Omega$.m$^2$), and can be considered a (differential) resistance. For the non-Faradaic (capacitive) process, the defining curve is very different, and it is a curve of charge, often defined per amount of electrode material, as function of electrode potential, see Fig.~\ref{fig_cartoon}f. The (inverse of the) derivative along the curve is the electrode capacitance, a property with dimension for instance F or F/g, which is a function of the charging degree. (In Fig.~\ref{fig_capacitance} we present data and theory for the capacitance of two electrode materials.)

Let us analyze what happens when the current approaches zero and the system goes to equilibrium. For the Faradaic process, for a very low current 
the electrode potential goes to a value determined by the Nernst equation. This potential depends on the activity of the reactant and product species in the bulk phases. With the activities (chemical potentials) of these species fixed, there is no way to 
modify the equilibrium electrode potential by pushing in charge. The charge would leak away by the electrode reaction, and ions diffuse out, and the system returns to the Nernst potential. Therefore, for a Faradaic process the well-known tables for (half-cell) standard electrode potentials apply. All of this is different in a non-Faradaic process. Here, an experimenter has control over the equilibrium electrode potential by injecting extra charge. This is the extra degree of freedom 
already referred to by Mohilner~\cite{Mohilner}. 
This extra parameter, or degree of freedom, is what distinguishes the non-Faradaic process from the Faradaic process.

Let us discuss the difference between Faradaic and non-Faradaic processes for one specific technology, Capacitive Deionization (CDI). CDI is a water desalination method which uses sets of porous electrodes that are charged and discharged in a cyclic manner, in this way adsorbing and desorbing ions in the electrodes from water, 
while electronic charge is also cyclically stored and released~\cite{Biesheuvel_Arxiv}. 
During charging of the cell, the electrode potential gradually changes. In some literature, this process of ion storage has sometimes been described as Faradaic when redox-active groups in the electrode were involved, also sometimes by authors of this document~\cite{Porada_PIMS, Suss_EES, He}. However, 
ion storage in CDI electrodes is always a non-Faradaic (capacitive) process, fitting in with the description of non-Faradaic processes as described above. In addition to ion storage, there can be a Faradaic process in CDI, leading to a (small) steady current on top of the capacitive current. In these electrode 
reactions, reactants and products come from, and go to, a bulk phase. An example is the splitting of water in oxygen and protons, or a reaction of \ce{O2} to \ce{H2O2}
~\cite{Dykstra2017,Zhang_Waite_WR_2018,Algurainy}.

\begin{figure}
\centering
\includegraphics[width=0.55\textwidth]{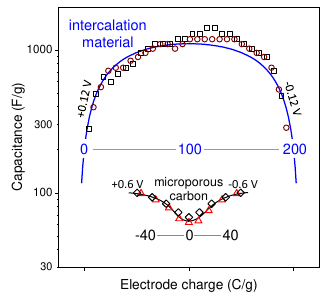}
\vspace{-1pt}
\caption{The capacitance of two types of porous electrode materials, 
the NiHCF intercalation material discussed in the main text, compared to data for microporous activated carbon. Preparation details for NiHCF in ref.~\cite{Porada_Smith}, measured in a three-electrode cell with a 3 M KCl Ag/AgCl electrode in 1 M \ce{Na2SO4}. Data for microporous carbon electrodes in 20 and 80 mM NaCl solution (red triangles and black diamonds, resp.), based on a two-electrode experiments at various charging voltages~\cite{Kim}.}
\label{fig_capacitance}
\end{figure}

\begin{figure}
\centering
\includegraphics[width=1\textwidth]{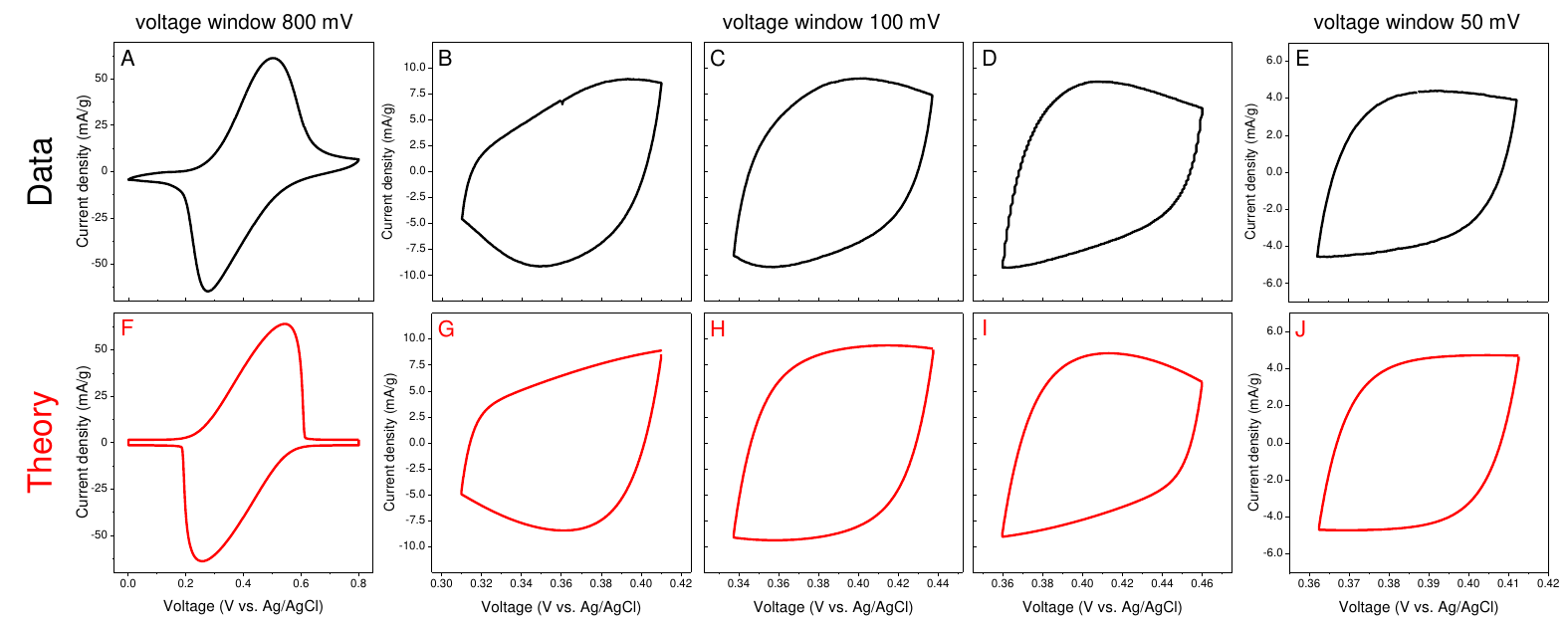}
\vspace{-10pt}
\caption{Cyclic Voltammetry (CV) diagrams for a NiHCF intercalation material tested in a three-electrode setup for the same conditions as described in Fig.~\ref{fig_capacitance}. Testing for different potential windows and voltage midpoints. Top row provides data, and bottom row theoretical predictions using a simple RC model. All theory curves only take as input the end-points of the potential window, 
the $g$-parameter relevant in the extended Frumkin isotherm, and a parameter $\mathcal{P}$ that is the product of scan rate, resistance, and maximum electrode charge, see main text.}
\label{fig_CV}
\end{figure}

An important type of electrode 
for CDI is the class of intercalation materials, for instance nickel hexacyanoferrate (NiHCF), a Prussian Blue analogue~\cite{Porada_Smith,Singh_PRA}. There is quite some discussion on whether ion storage in this material is by a Faradaic or a non-Faradaic (capacitive) mechanism. We argue that in line with the explanations provided above, and the literature sources cited, ion storage in an intercalation material is a non-Faradaic process. This position is also supported by the possibility to measure the 
capacitance of this electrode, see Fig.~\ref{fig_capacitance}, where we also compare with data for the capacitance of microporous activated carbon~\cite{Kim}.

The ability to measure the capacitance curve of NiHCF materials, in the same way that we can for microporous carbons, indicates that these materials are capacitive, and an electrode process involving this material is non-Faradaic. Nevertheless, it is sometimes argued that intercalation materials show features in a cyclic voltammetry (CV) experiment, namely peaks, that are evidence of a Faradaic process, while the CV diagrams for materials such as porous carbons 
do not have peaks and are more rectangular, and such diagrams point to these materials being capacitive. 
However, we will demonstrate that broad peaks obtained for intercalation materials in CV diagrams are due to their high capacitance when they are around halfway filled with cations, see Fig.~\ref{fig_capacitance}, while their capacitance drops off steeply at both high and low intercalation degrees, $\vartheta$, as described by the extended Frumkin isotherm. The dropping off of capacitance is because of the term $\ln\,\vartheta/\left(1-\vartheta\right)$ in the isotherm, see Eq.~(1) that we discuss below. Because of this term, approaching the limits of $\vartheta \rightarrow 0$ and $\vartheta\rightarrow 1$ leads to a faster and faster increase in electrode potential and thus a steeper decay in capacitance. Thus, when such a material is CV-tested in a large potential window, it will show a broad peak in each half cycle, see Fig.~\ref{fig_CV}, panels a and f, similar to data in Fig.~6 in ref.~\cite{Lumley}. This we see experimentally (panel a), and is in exact agreement with a theoretical RC model that combines an Ohmic resistance \textit{R} with a capacitance \textit{C} based on the extended Frumkin isotherm (a model for the capacitive behavior of the NiHCF electrode), see Fig.~\ref{fig_CV}f. Thus, there isn't much 
that is Faradaic about these broad peaks. Instead, they are simply the consequence of the capacitance not being constant. If we cycle 
this material in a smaller potential window, the curves become more and more rectangular, because in this smaller window the capacitance is more constant. 
Thus the broad peaks observed when we cycle NiHCF in a very wide potential window simply come from the fact that this intercalation material has the very interesting property that it has a maximum capacitance when filled with cations to around 50\%, and this capacitance decays when close to full occupancy, or close to empty. This is why this material --without there being any evidence of a Faradaic process-- shows 
broad peaks in a CV diagram when the potential window is chosen very wide, with the peaks gone when we cycle in a smaller potential window. These highly interesting features are not observed for a microporous carbon electrode, because these materials have a minimum in capacitance when uncharged, see Fig.~\ref{fig_capacitance}, and then at higher charge (both negative and positive), capacitance is relatively constant. 

We show in Fig.~\ref{fig_CV} results of calculations and experiments of a CV analysis for a NiHCF intercalation material as electrode (preparation details in ref.~\cite{Porada_Smith}). The electrode is around 450 $\mu$m thick, and per unit area has a mass of around 364 g/m$^2$ `active' NiHCF particles. The sample had an area of 20 cm$^2$, thus contained a mass $m_\text{el}=730$ mg of NiHCF material. 
As Fig.~\ref{fig_capacitance} shows, the maximum capacity per g of 
NiHCF that we can reach experimentally is around 200 C/g (i.e., the difference in electronic charge between the material loaded with cations and devoid of cations), 
and the maximum capacitance is around 1000~F/g. We show in Fig.~\ref{fig_CV} CV curves for three potential windows (PWs) and scanrates (SRs) (PW=800 mV \& SR=6 mV/min; PW=100 mV \& SR 0.6 mV/min; PW=50 mV \& SR=0.3 mV/min), and for the intermediate potential windows we use three values for the offset voltage (OV=360, 390 and 410 mV vs Ag/AgCl). As can immediately be observed, dependent on the window, we can have all possible shapes: we can have broad peaks, or the diagrams are more rounded, or they can even be very rectangular. We can also see that in all cases a theoretical calculation which uses a single resistance in combination with the extended Frumkin isotherm, which describes the relation between electrode voltage and electronic charge, 
fits data very well (details provided below). Intriguingly, in the widest window, the intercalation material shows the broad peaks that are considered to be an indication of a Faradaic electrode, but the same material, when cycled in a smaller window, shows a much more rectangular shaped CV diagram that is considered the fingerprint of a capacitive process. How is this possible? 

This is possible because this is a capacitive material, and 
the broad peaks are a consequence of the particular dependence of capacitance on charge for this material, 
and they do not relate to any purported Faradaic mechanism. 
Interestingly, using a simple theory that includes the extended Frumkin isotherm in a simple \textit{RC} network calculation --with the \textit{R} taken as a constant but not the \textit{C}, see Eq.~(1)-- we can make the theoretical curves match accurately to the experimental ones in all five cases considered in Fig.~\ref{fig_CV}. The theoretical curves reproduce the rectangular shape for the smallest potential window, the rounded shapes in the larger window, and the broad peaks in the widest window. The model also reproduces how the CV diagrams depend on the midpoint voltage, with the experimental sequence (panels b, c, d) accurately reproduced by the experimental diagrams (panels g, h, i). 
We argue that this result provides conclusive evidence that this type of intercalation material is capacitive. 

As discussed, the broad peaks in a CV diagram of NiHCF are the consequence of the decreasing capacitance of this material at the `edges' of its charging curve. This is different for other capacitive materials where capacitance is constant when the charge is taken to either very positive or very negative, such as is the case for microporous carbons, see Fig.~\ref{fig_capacitance}. Thus, the only feature that is unfamiliar about the intercalation materials, in contrast to materials such as microporous carbons, 
is that for NiHCF we have a curve for capacitance that is at a maximum in the middle of its charging range, and then drops off to the sides. 
CV diagrams for materials with a more constant capacitance (or with a minimum at intermediate charge) 
will be more `rectangular'. 

The RC network calculation is based on a very small number of simple equations. We described the electrode voltage of the intercalation material using the extended Frumkin isotherm,
\begin{equation}
{V_\text{cap}} = V_\text{ref} +  V_\text{T} \cdot \left(\ln \frac{c_\infty}{c_\text{ref}} - \ln \frac{\vartheta}{1-\vartheta}  - g' \, \left(\vartheta-\nicefrac{1}{2}\right)\right)
\end{equation}
and combine with a resistance $R$ in series with it. In Eq.~(1), $g$ is an ion-ion attraction parameter that we measured to be $g' \sim -3.5$~\cite{Porada_Smith}, while $V_\text{T}=RT/F$ is the thermal voltage, around 25.6 mV. We can fit the data for capacitance in Fig.~\ref{fig_capacitance} using Eq.~(1) based on the conversion $\Sigma = \Sigma_\text{max} \cdot \vartheta$, with $\Sigma_\text{max}=210$~C/g and $C=- \partial \Sigma / \partial V_\text{cap} = \Sigma_\text{max} / V_\text{T} \cdot   \left(\vartheta^{-1} + \left(1-\vartheta\right)^{-1}  + g' \right)^{-1} $. 
In the CV-analysis, Fig.~\ref{fig_CV}, we use the same relation between charge $\Sigma$ and intercalation degree $\vartheta$. 
Concentration $c_\infty$ is the \ce{Na+}-concentration in solution, and $c_\text{ref}$ is a reference concentration. 
In our calculations we use $V_\text{ref}^*=387.5$~mV for $V_\text{ref}+V_\text{T}\,\ln\left(c_\infty/c_\text{ref}\right)$. Thus, cycles that have $V_\text{ref}^*$ as their midpoints (such as panels f, h, j in Fig.~\ref{fig_CV}) have a symmetric shape. The voltage signal imposed is the sum of the voltage over the resistance, $V_\text{res}=I\cdot R$, and that over the electrode, 
$V_\text{cap}$. The change of electrode charge $\Sigma$ with time $t$ equals the current, $I$. 
Intriguingly, the entire problem can be formulated with only two parameters, the $g'$-factor in Eq.~(1), and a dimensionless factor $\mathcal{P}$ which is the product of scanrate (SR, in V/s), resistance ($\Omega$), the maximum charge $\Sigma_\text{max}$ (C/g), the electrode mass, $m_\text{el}$ (g), 
and the inverse of $V_\text{T}^2$. Fitted values for $\mathcal{P}$ in Fig.~\ref{fig_CV} as function of voltage window are: PW=800 mV: $\mathcal{P}=20$; 100 mV: $\mathcal{P}=4$; 50 mV: $\mathcal{P}=2$. Because $\mathcal{P}$ is expected to be proportional with SR, the $\mathcal{P}$-value for PW=800 mV is too small relative to the other two data sets. We reduced the value of $\mathcal{P}$ for PW=800 mV to make the peaks come closer. 

In conclusion, almost 70 years after David Grahame's discussion of the distinction between Faradaic and non-Faradaic electrode processes, his analysis is still of great importance. The most useful approach is to take the experimental or observational perspective, based on data of the response of an electrode to an ongoing current supply. For intercalation materials and other redox-active materials, different types of cyclic voltammograms are possible, and we provide 
one example that demonstrates that an RC model which implements a realistic model for electrode capacitance, can reproduce these various shapes accurately. Peaks in such diagrams do not 
imply a Faradaic mechanism. 

\newpage


\end{document}